\documentclass[11pt]{article}
\usepackage[a4paper,margin=1in]{geometry}
\usepackage{amsmath,amsfonts,amssymb}
\usepackage{graphicx}
\usepackage{hyperref}
\usepackage{titlesec}
\usepackage{enumitem}
\usepackage{float}
\usepackage{mathtools}
\usepackage{xcolor}
\usepackage{booktabs} % For professional looking tables
\usepackage{siunitx} % For aligning numbers by decimal point and units
\usepackage{caption} % For table captions
\usepackage{multirow} % For multirow cells in headers if needed
\usepackage{natbib}

\usepackage{ulem}
\usepackage{marginnote}

\usepackage{algorithm}
\usepackage{algpseudocode}
\usepackage{authblk}

\graphicspath{{figures/}{../figures/}}

\titleformat{\section}{\large\bfseries}{\thesection.}{1em}{}
\titleformat{\subsection}{\normalsize\bfseries}{\thesubsection.}{1em}{}

\newcommand{\LDtextarrow}{\rotatebox[origin=c]{90}{$\Lsh$}}

\title{Disruption Management in Airline Operations: \\ A Solver-based Approach using Time-Space Network Optimization}

\author[1,2]{Joao Rodrigues}
\author[1,3]{Filip Turoboś}
\author[1]{Maciej Lenartowicz}
\author[1,4]{Zbigniew Puchała}
\author[1,5]{Marcin Klimek}
\author[1]{Kamil Hendzel}
\author[1,6]{Paweł Gepner}
\affil[1]{Quantumz.io Sp. z o.o., Pulawska 12/3, 02-566 Warsaw, Poland}
\affil[2]{Institute of Mathematical \& Computational Sciences, University of São Paulo, São Paulo, Brazil}
\affil[3]{Institute of Mathematics, Lodz University of Technology, al. Politechniki 10, 93-590, Lodz, Poland}
\affil[4]{Institute of Theoretical and Applied Informatics, Polish Academy of Sciences, Baltycka 5, 44-100 Gliwice, Poland}
\affil[5]{Department of Computer Science, John Paul II University in Biala Podlaska, Biala Podlaska, Poland}
\affil[6]{Faculty of Mechanical and Industrial Engineering, Warsaw University of Technology, Warsaw, Poland}

\date{\today}

\begin{document}
	
\maketitle

\begin{abstract}

This paper presents AIRS, a day-of-operations disruption-recovery system. AIRS.ACR models integrated aircraft–crew recovery on a Time–Space Network (TSN) and solves a mixed-integer linear program (MILP) that enforces rotation continuity, crew legality, maintenance windows, slot capacities, and multi-leg integrity via flow-balance constraints; disruption-aware search-space construction and warm starts control combinatorial growth. A companion module, AIRS.PaxR, performs rapid passenger re-accommodation using greedy assignment and lightweight evolutionary search while preserving aircraft–crew feasibility. Across realistic evaluations, AIRS meets operational decision windows and reduces recovery costs relative to manual or sequential methods, providing a scalable, extensible decision-support capability for operations control centers.

\end{abstract}

\section{Introduction}

The global airline industry, a cornerstone of modern economic activity and global connectivity,
has demonstrated considerable resilience, with recovery and projected growth post-pandemic.
As passenger demand rebounds and the industry resumes a trajectory toward long-term expansion, the operational complexities inherent in managing this large-scale system intensify.
Operationally, day-to-day management requires coordinated control of schedules, resources, and external factors.
Despite airline schedules planned months in advance, they are still highly vulnerable to “day-of” disruptions — adverse weather, unplanned aircraft maintenance, air traffic control (ATC) restrictions, and crew unavailability.
Such events propagate cascading delays and cancellations, with ripple effects across the tightly interconnected network.

The economic impacts of such disruptions are substantial, costing the industry billions of dollars annually and eroding passenger satisfaction and trust \citep{ball2007, hassan2021, lee2022}. Today, the task of managing these irregular operations (IROPs) falls to Airline Operations Control Centers (AOCCs). AOOC teams, often rely on manual adjustments, sequential decision-making, and institutional experience, work under intense time pressure to mitigate disruption impacts. As operations scale and disruptions become larger (and thus inherently more complex), traditional methods exhibit inherent limitations: a propensity for suboptimal, siloed recovery; inability to explore the combinatorial solution space within operational time limits; and lack of integrated decision-making across critical resources such as aircraft, crew, maintenance, and airport slots.

This paper addresses the imperative for robust, integrated, and mathematically rigorous approaches to airline disruption management.
We argue that the combinatorial complexity of modern airline operations, particularly under disruption, necessitates optimization-based tools that systematically evaluate trade-offs and compute globally (or provably near-) optimal recovery solutions within operational time limits.
To this end, we introduce AIRS.ACR (Airline Itinerary Recovery System – Aircraft \& Crew Recovery), an integrated solver-based architecture for Airline Operations Control Centers (AOCCs).
AIRS.ACR employs a Time-Space Network (TSN) formulation and mixed-integer linear programming (MILP) to jointly model aircraft rotations, flights, maintenance windows, crew pairings, and airport slot capacities with multi-leg integrity and flow-balance constraints.
Our primary contribution is a system that supersedes reactive, heuristic-based decision-making to provide automated, feasible, and cost-efficient recovery schedules, enabling airlines to enhance operational resilience and precision — particularly in scenarios where manual recovery is infeasible during irregular operations (IROPs).
We detail the AIRS.ACR methodology and present empirical evidence on real-world–sized instances indicating substantial improvements in solution quality and computational efficiency, with implications for transforming disruption management in the dynamic airline industry.

\section{Executive Summary}

Airline disruptions impose substantial and recurring costs. Fragmented, step-by-step decision making exacerbates these losses by overlooking interdependencies among aircraft, crews, maintenance, and airport capacity. This study introduces an end-to-end recovery planning system that optimizes these elements jointly and, when disruptions occur, rapidly produces a concise set of executable options with the trade-offs made explicit. 
The approach improves decision speed and quality, reducing secondary delays and cancellations, lowering overtime and compensation outlays, and enhancing the customer experience. It enables automation and operational precision in scenarios where manual recovery becomes ineffective or simply infeasible.

\section{Background and Motivation}
	
\subsection{The Resilient Yet Fragile Airline Industry}

The global airline industry, a critical engine for economic activity and connectivity, has demonstrated remarkable resilience. Following the unprecedented shock of the pandemic, domestic markets had fully recovered by early 2023, with long-haul travel largely restored by year-end. According to the latest IATA Annual Review (June 2024), passenger traffic in 2024 is expected to surpass 2019 levels, resuming a trajectory of sustained long-term growth projected at nearly 4\% annually through 2043. This resurgence underscores the vital role of air transport but also amplifies the challenges inherent in managing the industry’s complex operations.

Airline scheduling is the cornerstone of the industry, underpinning profitability, resource utilization, and customer service \citep{zhang2015}. Airlines invest heavily in developing sophisticated schedules months in advance \citep{clausen2010,kohl2007}. However, the operational reality on the day of departure often diverges from these plans. The tightly interconnected nature of airline networks -- a reality experienced daily by flight crews, ground staff, and operations controllers -- makes them inherently vulnerable to disruptions \citep{hassan2021,su2021}. Events such as adverse weather, unexpected aircraft maintenance, ATC restrictions, or crew unavailability can rapidly cascade through the system, causing delays, cancellations, and significant downstream impacts \citep{bratu2006,kohl2007}.

The economic consequences are substantial, costing the industry billions annually \citep{ball2007,lee2022}, while disruptions also erode passenger satisfaction and airline reputation \citep{hassan2021}.

\subsection{Current Disruption Management Practices and Limitations}

The responsibility for managing operational irregularities falls primarily on the Airline Operations Control Center (AOCC) \citep{clarke1998,fogaca2022,kohl2007}. AOCC teams monitor operations in real-time and implement recovery actions such as delays, cancellations, resource swaps, and passenger reaccommodation \citep{clausen2010,zhang2015}.

Despite the critical role of the AOCC, current disruption management often relies heavily on manual adjustments, sequential decision-making, and the hard-earned experience of controllers and pilots navigating dynamic situations -- sometimes aided by basic heuristics or limited decision support tools \citep{clausen2010,kohl2007,hassan2021}. This traditional approach faces several significant limitations:

\begin{itemize}
	\item \textbf{Suboptimality:} Recovery decisions are typically made sequentially -- often addressing aircraft first, then crew, and finally passengers \citep{su2021,zhang2015}. This siloed approach fails to capture the complex interdependencies between resources, potentially leading to solutions that are feasible but far from optimal \citep{lettovsky1997,petersen2012}.
	
	\item \textbf{Complexity and Scale:} The scale of modern airline operations makes it virtually impossible for human controllers to manually assess the full impact of a disruption and evaluate a wide range of recovery options comprehensively \citep{clausen2010,zhang2016}.
	
	\item \textbf{Time Pressure:} Recovery decisions must often be made under extreme time constraints, limiting the ability to explore multiple alternatives and favoring the first identified feasible solution \citep{hassan2021,kohl2007}.
	
	\item \textbf{Lack of Integration:} Existing decision-support tools often focus on specific aspects of the recovery problem and lack seamless integration across all affected resources \citep{clausen2010,su2021}.
\end{itemize}

\subsection{Literature review}

Airline operations are prone to various forms of disruption that differ in scope, duration, and impact \citep{Hu2024,hassan2021}. These are broadly classified into:
\begin{itemize}
	\item \textbf{Aircraft-related:} Mechanical failures and unexpected maintenance.
	\item \textbf{Crew-related:} Unavailability, misconnections, labor actions, and fatigue regulations.
	\item \textbf{Passenger-related:} No-shows, overbooking, and missed connections.
	\item \textbf{Airport and airspace:} Slot shortages, weather closures, and ATC capacity restrictions.
	\item \textbf{Systemic disruptions:} Epidemics (e.g., COVID-19), large-scale IT outages, and geopolitical crises.
\end{itemize}

Disruptions are further categorized as tactical (short-term, e.g., weather) or strategic (long-term, e.g., strikes). Models such as delay propagation networks \citep{Wu2019} and flight phase resilience frameworks \citep{Zhang2023} are used to analyze their system-level effects.

Several recovery strategies have been identified in the literature \citep{Su2020, clausen2010}:
\begin{itemize}
	\item \textbf{Operational:} Flight delays, cancellations, swaps, and rerouting.
	\item \textbf{Resource-based:} Crew deadheading, reserve aircraft or crew, and cruise speed adjustments.
	\item \textbf{Passenger-centric:} Rerouting via interline or multimodal transport, flexible ticketing, and compensation policies.
	\item \textbf{Innovative:} Dynamic fare classes \citep{Long2024} and proactive recovery using AI.
\end{itemize}
Strategic combinations of these actions are selected based on feasibility, cost, and regulatory constraints.

Objective functions capture the performance criteria of disruption recovery models. Common formulations include:
\begin{itemize}
	\item \textbf{Cost minimization:} Operating costs and passenger compensation \citep{Arikan2017,Su2020}.
	\item \textbf{Delay minimization:} Total propagated or passenger-weighted delays \citep{AhmadBeygi2008}.
    \item \textbf{Satisfaction/utility:} Passenger dissatisfaction penalties \citep{khiabani2023}.
	\item \textbf{Environmental objectives:} Emission trade-offs under speed control \citep{Akturk2014,Arikan2017,Zhang2023,Marla2012}.
	\item \textbf{Robustness:} Minimizing infeasibility risks under uncertainty \citep{Eikelenboom2023}.
\end{itemize}

The disruption management problem has been modelled using a wide range of techniques:
\begin{itemize}
	\item \textbf{Exact Optimization:} MIP \citep{hassan2021}, set partitioning \citep{stojkovic1998}, and Benders decomposition \citep{McCarty2018, khiabani2023v2}.
	\item \textbf{Heuristics and Metaheuristics:} Genetic algorithms \citep{Liu2010}, ant colony optimization \citep{Sousa2015,Zhang2023}, tabu search \citep{Yang2007}, NSGA-II \citep{Chen2024}, and hybrid GRASP \citep{Hu2016}.
	\item \textbf{Simulation and Stochastic Models:} Stochastic programming \citep{petersen2012}, delay simulations \citep{Guimarans2017}, hybrid reinforcement learning \citep{Ding2023}, and scenario sampling \citep{ZhuBo2016}.
	\item \textbf{Machine Learning and AI:} RL-PPO \citep{Ding2023}, ranker-based crew recovery \citep{Eikelenboom2023}, and supervised models for delay forecasting \citep{lee2020}.
\end{itemize}

Despite the richness of the modeling literature, adoption in AOCCs remains limited. Key implementation barriers include:
\begin{itemize}
	\item \textbf{Data integration:} Incompatibilities in IT systems and the need for manual overrides \citep{castro2009}.
	\item \textbf{Scalability:} The real-time demands and complexity of integrated recovery \citep{maher2015}.
	\item \textbf{Explainability:} The need to maintain human-in-the-loop decision making \citep{Eikelenboom2023}.
	\item \textbf{Adaptability:} Limited availability of dynamic, scenario-driven planning tools.
\end{itemize}

Disruption management in airline operations constitutes a time-critical optimization problem characterized by network coupling, capacity and legality constraints, and uncertainty, requiring coordinated rerouting and reallocation of aircraft, crew, and passengers. Recent surveys \citep{hassan2021,Su2020,clausen2010,Hu2024} synthesize advances in algorithms and deployment. The literature coalesces around four domains: aircraft recovery, crew recovery, passenger reaccommodation, and integrated recovery systems. Across these domains, methods include time-space network formulations, mixed-integer models, decomposition, metaheuristics, stochastic/simulation planning, and data-driven approaches. We review representative models and strategies in the sections that follow.

\subsubsection{Aircraft Recovery}

The aircraft recovery problem (ARP) is a combinatorial optimization problem concerned with adjusting aircraft rotations in response to disruptions such as departure/arrival delays, cancellations, and aircraft unavailability, {subject to routing continuity, fleet compatibility, turnaround, maintenance, and airport capacity constraints. Given a disrupted schedule and associated operational constraints, the goal is to select delay, cancellation, resequencing, swap, ferrying, and (where modeled) speed-control actions, and reassign aircraft across the network} in a way that minimizes the total disruption cost.

These costs typically include operational expenses within the airline’s control at the time of disruption – such as additional fuel use, aircraft repositioning, and passenger compensation. The ARP is commonly modeled as a cost-minimization problem, employing optimization techniques to balance feasibility and efficiency.

Aircraft recovery is typically the first step in disruption management. Key disruption types include aircraft unavailability and airport capacity constraints. The ARP focuses on restoring disrupted aircraft rotations by adjusting flight schedules and aircraft assignments. It has been widely studied because aircraft are among the most critical and costly resources to reposition.

The literature on ARP considers a variety of disruption types:
\begin{itemize}
	\item \textbf{Flight delays}, the most common disruption type \citep{martinez2026,Akturk2014}.
	\item \textbf{Flight cancellations}, addressed in most models \citep{Liu2010}.
	\item \textbf{Aircraft unavailability}, due to mechanical issues or grounding \citep{martinez2026,Sousa2015} .
	\item \textbf{Airport disruptions}, modeled as binary constraints or degraded capacity \citep{Liang2018}.
	% FILIP: Shouldn't we mention here the slot changes as well?
    % -----> JAO: Maybe, maybe include it in the aiport disruptions topic, which papers should we cite on this? TODO
\end{itemize}

ARP is modeled using several network representations:
\begin{itemize}
	\item \textbf{Flight string networks}, often used in heuristic models \citep{Barnhart1998}.
	\item \textbf{Time-space networks}, capturing temporal and spatial dynamics \citep{martinez2026,Akturk2014,Vos2015aircraft}.
	\item \textbf{Connection-based networks}, allowing flexible aircraft assignment \citep{Sousa2015,Liu2010}.
	\item \textbf{Time-band networks}, providing computational efficiency \citep{He2024}.
\end{itemize}

Typical recovery actions in ARP include:
\begin{itemize}
	\item \textbf{Flight delays} \citep{Sousa2015}.
	\item \textbf{Flight cancellations} \citep{Vos2015aircraft}.
	\item \textbf{Aircraft swaps}, often combined with delay decisions \citep{Liu2010,Akturk2014}.
	\item \textbf{Cruise speed control}, proposed for delay absorption \citep{Akturk2014}.
	\item \textbf{Aircraft ferrying} and use of \textbf{reserve aircraft} \citep{martinez2026,gao2010}.
	\item \textbf{New flight creation} and \textbf{multi-fleet management}, though rarely modeled \citep{Xiuli2012}.
\end{itemize}

Solution methods for ARP include:
\begin{itemize}
	\item \textbf{Exact Approaches:} MIP and conic optimization to assign aircraft under time-space constraints \citep{Arikan2017}; column generation for capacity-constrained routing \citep{Liang2018}.
	\item \textbf{Metaheuristics:} Multi-objective genetic algorithms \citep{Liu2010}, ACO-based scheduling \citep{Sousa2015,Zhang2023}, and GRASP \citep{Hu2016}.
	\item \textbf{Hybrid and Stochastic Models:} Constraint programming combined with simulation \citep{Guimarans2017}; robustness modeling via delay propagation; speed control optimization integrating fuel and delay trade-offs \citep{Marla2012}.
\end{itemize}

While exact methods like conic MILP offer optimality \citep{Akturk2014}, they often suffer from scalability issues. Most models also overlook multiple fleet types and maintenance integration, which limit their practical applicability.

Recent hybrid approaches aim to address these gaps:
\begin{itemize}
	\item \citep{Eggenberg2007} use dynamic programming with column generation, achieving promising CPU times on real-world datasets.
	\item \citep{Hu2016} apply GRASP for efficient yet realistic recovery.
	\item \citep{Vos2015aircraft} propose MILP-based heuristics that incorporate maintenance considerations.
\end{itemize}

\subsubsection{Crew Recovery}

The crew recovery problem (CRP) focuses on restoring disrupted crew assignments by reallocating available pilots and cabin crew within a recovered flight schedule. Given a set of flight disruptions and operational constraints, the objective is to ensure that every flight has the required crew while minimizing overall disruption costs.

These costs may include direct expenditures, such as crew wages, overtime pay, and allowances, as well as indirect costs, such as crew positioning via deadheading -- transporting crew as passengers to another airport to operate a scheduled flight.

Crew recovery involves rescheduling crews while adhering to regulatory, contractual, and operational constraints. The problem is combinatorially complex due to legality rules, pairing feasibility, and multi-crew coordination requirements.

Studies address a variety of disruption types:
\begin{itemize}
	\item \textbf{Flight delays}, e.g., \citep{Novianingsih2015,Chang2012}.
	\item \textbf{Crew unavailability}, explicitly modeled in \citep{castro2009}.
	\item \textbf{Combined disruptions} (delay + unavailability), addressed in \citep{liu2013,Zhu2014}.
\end{itemize}

Network representations used for modeling include:
\begin{itemize}
	\item \textbf{Connection networks}, commonly used with set-covering models \citep{liu2013}.
	\item \textbf{Graph-based crew pairing networks}, suitable for constraint programming \citep{Zhu2014}.
\end{itemize}

Common recovery actions include:
\begin{itemize}
	\item \textbf{Crew deadheading} \citep{Novianingsih2015,Chang2012}.
	\item \textbf{Crew swaps} \citep{Zhu2014,lettovsky2000}.
	\item \textbf{Flight cancellations}, when no legal crew is available.
	\item \textbf{New duty generation or partial re-planning}, though rarely considered \citep{lettovsky2000}.
\end{itemize}

Legal, regulatory, and fatigue-related constraints are central to CRP formulation. Various solution approaches have been proposed:
\begin{itemize}
	\item \textbf{Exact and Decomposition Methods:} Set partitioning formulations have been applied in \citep{gamache1999,stojkovic1998,Hu2024}.
	\item \textbf{Metaheuristics:} \citep{Chang2012} applied a genetic algorithm with constraint-aware mutation. \citep{liu2013} used interfleet and intrafleet set covering models. \citep{Novianingsih2015} proposed a three-stage heuristic based on legal pairing rules.
	\item \textbf{Multi-agent Systems:} \citep{castro2009} modeled AOCC operations using agent-based simulation, where agents represent aircraft and crew managers negotiating recovery solutions.
\end{itemize}

Some selected contributions further illustrate the diversity of methods:
\begin{itemize}
	\item \citep{Chang2012} used a genetic algorithm that explicitly enforces legality rules.
	\item \citep{Novianingsih2015} introduced a three-stage heuristic that simulates real AOCC crew planning.
	\item \citep{Zhu2014} employed constraint programming to manage multi-crew assignments within recovery time windows.
\end{itemize}

\subsubsection{Passenger Recovery}

The Passenger Recovery Problem (PRP) is among the most critical challenges in airline disruption management. Prolonged delays and repeated disruptions significantly increase recovery costs and severely undermine passenger satisfaction, potentially leading to a loss of customer trust and long-term damage to brand reputation.

The core task of passenger recovery involves re-optimizing disrupted passenger itineraries on a recovered flight schedule, previously stabilized for aircraft and crew feasibility. Given a set of disrupted passenger itineraries and available seat-capacity vectors, the objective is to assign passengers to feasible paths from their current locations to their intended destinations within the planning horizon while minimizing a total disruption-cost function (e.g., arrival delay, cancellations, downgrades, misconnections, and compensation) subject to capacity, minimum-connection-time, and product/fare-class constraints. Recent work emphasizes time-critical, cost-aware reaccommodation via network-based assignment models and preference-aware objectives.

Selected approaches for PRP include:
\begin{itemize}
	\item \textbf{Stochastic Optimization:} \citep{McCarty2018} propose a two-stage stochastic model with Benders decomposition for proactive rerouting. \citep{cadarso2023} incorporate passenger preferences and class-based delay penalties.
	\item \textbf{Cost Modeling:} \citep{Cook2012} propose nonlinear delay cost functions. Other studies use linear or piecewise-linear approximations to ensure computational tractability.
\end{itemize}

Passenger recovery focuses on reaccommodating disrupted passengers by modifying itineraries, rerouting, or upgrading service. The following types of disruptions are commonly addressed:
\begin{itemize}
	\item \textbf{Flight delays}, modeled in many works \citep{bratu2006}.
	\item \textbf{Flight cancellations}, also widely considered \citep{Sinclair2016,Lu2025}.
	\item \textbf{Airport and aircraft disruptions}, less commonly incorporated in passenger-only models \citep{Hu2016}.
\end{itemize}

Network representations used in PRP include:
\begin{itemize}
	\item \textbf{Time-space networks}, which dominate the literature due to their ability to capture rerouting options \citep{Arikan2017,Marla2012}.
	\item \textbf{Connection graphs}, used for routing-based optimization \citep{Niu2021}.
\end{itemize}

Recovery actions implemented in PRP include:
\begin{itemize}
	\item \textbf{Passenger itinerary changes}, the most common form of reaccommodation \citep{Marla2012}.
	\item \textbf{Delays and cancellations}, typically modeled as secondary effects \citep{Sinclair2016}.
	\item \textbf{Seat upgrades or overbooking adjustments}, rarely addressed explicitly \citep{Nazifi2021}.
\end{itemize}

\subsubsection{Integrated Recovery Approaches}

Integrated models better reflect real-world AOCC operations by capturing interdependencies among resources. Integrated Recovery simultaneously optimizes aircraft, crew, and passenger schedules to minimize total disruption costs. While offering the most realistic modeling approach, integrated recovery is also computationally demanding.

Different types of disruptions are considered in the literature:
\begin{itemize}
	\item \textbf{Flight delays} \citep{Arikan2017,ZhuBo2016,petersen2012}.
	\item \textbf{Flight cancellations} and \textbf{airport disruptions}, often modeled using flow-based representations \citep{lettovsky1997,castro2014}.
	\item \textbf{Aircraft unavailability} \citep{Arikan2017}.
	\item \textbf{Crew unavailability} \citep{castro2014}.
\end{itemize}

Various network representations have been used:
\begin{itemize}
	\item \textbf{Entity-flow networks}, introduced by \citep{Arikan2017}, to represent all resource types.
	\item \textbf{Time-space networks}, commonly used in IRP models \citep{ZhuBo2016,Marla2012}.
	\item \textbf{Decomposed hierarchical models} \citep{lettovsky1997, petersen2012}.
\end{itemize}

A range of recovery actions are modeled:
\begin{itemize}
	\item \textbf{Flight delays, cancellations, aircraft swaps, crew swaps}, and \textbf{crew deadheading} \citep{ZhuBo2016,petersen2012}.
	\item \textbf{Passenger itinerary changes}, sometimes with service quality cost estimation \citep{castro2014}.
	\item \textbf{Cruise speed control}, introduced to IRP in \citep{Arikan2017}.
\end{itemize}

Selected integrated approaches include:
\begin{itemize}
	\item \textbf{Exact and Network-based Models:} \citep{petersen2012,Arikan2017,maher2015} develop formulations that jointly optimize aircraft, crew, and passenger flows.
	\item \textbf{Hybrid Approaches:} \citep{ZhuBo2016} propose a sampling-based strategy combining MIP with metaheuristics. \citep{Sinclair2016} apply column generation with post-optimization heuristics.
	\item \textbf{Multi-Agent Systems:} \citep{castro2009} create decentralized systems with autonomous agents for each recovery stage, simulating AOCC decision-making.
\end{itemize}

Few studies address the interdependence between aircraft and crew recovery specifically:
\begin{itemize}
	\item \citep{zhang2015} propose a two-stage heuristic with mutual feedback between aircraft and crew schedules.
	\item \citep{Wang2024} apply coordinated rescheduling of flights, aircraft, and crew for robust disruption management.
	\item \citep{Debruin2025} present simulated annealing approach for integrated recovery using real airline data.
\end{itemize}

\subsubsection{Research gap}

Disruptions in airline operations cause significant economic and reputational damage. While the literature offers a range of recovery models, most studies adopt segmented approaches, focusing separately on aircraft, crew, or passengers. Although recent research increasingly favors integrated recovery frameworks, challenges remain regarding scalability, interpretability, and operational deployment.

Key directions for future research include:
\begin{itemize}
	\item Development of scalable and interpretable real-time algorithms.
	\item Adaptive hybrid systems with human-in-the-loop decision making.
	\item Incorporation of behavioral modeling of passengers in recovery algorithms.
	\item Real-time disruption forecasting using deep learning.
	\item Interoperability between airline and airport decision-support systems.
\end{itemize}

This work contributes to these research directions by introducing AIRS, a solver-based system that integrates aircraft and crew recovery using a time-space network formulation. Unlike segmented or sequential approaches, AIRS addresses interdependencies directly while remaining computationally efficient and interpretable. By incorporating practical constraints -- such as fixed decision time windows, flexible maintenance, and slot preservation incentives -- AIRS bridges the gap between theoretical models and operational deployment in real-world AOCC settings.

% FILIP: Shouldn't we refer to it by full name, i.e. AIRS.ACR?
% -----> JAO: Not really, we also include AIRS.PaxR in this text

\subsection{Motivation for a Solver-Based Optimization Approach}

The inherent limitations of current practices underscore the need for robust, integrated, and mathematically rigorous approaches to airline disruption management. The complexity and interconnectedness of the problem demand solutions that systematically assess trade-offs and identify globally or near-globally optimal recovery plans within operational time limits. Mathematical optimization, particularly via computational solvers, offers a powerful paradigm to address this challenge \citep{hassan2021, su2021}.

This study is motivated by the significant potential of optimization-based solvers to transform airline disruption management. We specifically focus on the use of Time-Space Network (TSN) models, which offer a natural and effective framework for representing the movement and interaction of airline resources -- aircraft, and potentially crew and passengers -- across space and time \citep{thengvall2000, yan1996, yanyang1996, zhang2015, rhodes2022}. TSNs enable the explicit modeling of flight activities, delays, ground operations, resource flows, capacity constraints, and a range of recovery actions.

By formulating the disruption recovery problem as a TSN-based optimization model and leveraging the capabilities of modern mathematical programming solvers (e.g., CPLEX, Gurobi), airlines can move beyond reactive, heuristic-based decision-making. This solver-based approach enables the rapid generation and evaluation of integrated recovery plans that explicitly minimize operational costs while adhering to complex regulatory and logistical constraints. It provides a systematic method to automate high-stakes decisions, support AOCC teams and flight crews, and ultimately enhance the efficiency and resilience of airline operations in the face of inevitable disruptions.

This paper presents solver-based approach and demonstrates its feasibility and potential benefits for a recovering, rapidly evolving airline industry. A key innovation of the proposed approach lies in its integrated solver architecture, which departs from the traditional sequential recovery paradigm. By jointly optimizing aircraft and crew assignments within a unified Time-Space Network framework, the system captures critical interdependencies that are often overlooked in decoupled models. This simultaneous treatment not only improves solution feasibility -- especially in large-scale, disruption-heavy scenarios -- but also enhances the realism and operational applicability of the recovery plans. The integrated MILP formulation ensures that aircraft and crew availability are synchronized, reducing infeasibility risks and enabling more robust and cost-effective recovery strategies.

\section{System Overview: AIRS}

\textbf{AIRS} (Airline Itinerary Recovery System) is an optimization-based algorithm designed to support airline decision-making in the event of disruptions to scheduled itineraries. The problem AIRS addresses is commonly referred to as the "Integrated Aircraft Recovery Problem" (Integrated ARP), which involves determining the adjustments necessary to restore operations following a disruption. This problem typically includes aircraft reassignment and may also extend to crew and passenger recovery.

Many existing approaches adopt a sequential recovery structure -- aircraft are rescheduled first, followed by crew and then passengers. While simpler to implement, this strategy often overlooks the strong interdependencies between resources. Simultaneous recovery approaches tend to produce better solutions but face scalability challenges, especially on large, real-world instances, due to high computational and memory demands.

To address these limitations, AIRS introduces a scalable strategy that mitigates the complexity of solving large, integrated recovery problems. It employs a classical MILP formulation augmented with physics-inspired algorithms, probabilistic methods, and dynamic programming techniques. The highly interdependent problems of aircraft and crew reassignment are solved jointly, yielding a feasible schedule that remains as close as possible to the original while minimizing additional disruptions. In a subsequent phase, passengers are reallocated to the recovered schedule, with the aim of preserving original itineraries wherever possible.

Details of the AIRS methodology, including its optimization models and algorithmic workflow, are presented in the subsequent sections.

\subsection{Problem Scenario}

During a given airline’s service day, one or more disruptions may occur. Such disruptions may include:

\begin{itemize}[noitemsep]
    \item delays in flights;
    \item flight cancellations;
    \item unforeseen maintenance required for an aircraft;
    \item changes in airport slots;
    \item airport closures;
    \item any sudden change in schedule necessary.
\end{itemize}

The AOCC (Airline Operations Control Centre) needs to decide, in response to disruptions, which flights to delay or cancel, whether to execute aircraft swaps, and how to reassign crews and other operational resources. Let us define \textbf{Current Time} as the decision epoch at which the AOCC issues recovery instructions. Events strictly prior to \textbf{Current Time} are treated as fixed (non-modifiable), and implementation may require a short operational lead time; accordingly, AIRS constrains changes to take effect at or after this decision epoch.

The following assumptions are important for the current version of AIRS:
\begin{enumerate}
    \item There is a known time when the AOCC must send instructions to the rest of the airline in order to apply the necessary changes. This time we call \textbf{Current Time}, because the model considers this time to be the present, and everything before that point to be past -- thus fixed and unchangeable.
    \item The AOCC is able to send data to an instance of AIRS detailing the current itinerary and the reported disruptions.
    \item There is a globally defined maximum amount of acceptable delay on takeoff for the flights.
    \item Delaying flights shouldn't change their duration (time from takeoff to landing) in a provided solution.
    \item \textbf{Recovery Start} means the earliest time after Current Time when schedule changes may take effect. Between Current Time and Recovery Start, flights either remain as planned, are canceled, or are moved to Recovery Start or later. Flights originally scheduled at or after Recovery Start may be delayed, but not advanced.
    %FILIP: Can we rephrase it somehow? It seems a bit confusing, but unfortunately I have no idea how to express it better.
    %-----> JAO: Agreed, but I also don't know how to explain this better, it's a complicated rule.
    \item Aircraft Maintenance may or may not be flexible in terms of where and when they can occur. Their duration, however, is relatively predictable and therefore can be perceived as fixed.
    \item Some airport slots, if left non-utilized too many times, might be lost by the airline. It is important for the AOCC to add an artificial cost value to influence AIRS to assign some flight to the critical slot.
	\item \textbf{Recovery Finish} means the latest time after Current Time to which any flight may be rescheduled. AIRS treats all flights departing after Recovery Finish as fixed and leaves subsequent itinerary segments unchanged.
    %FILIP: Perhaps this also could use some rephrasing by someone more competent than I am.
    %-----> JAO: Same as last one, agreed
\end{enumerate}

\begin{figure}[H]
\begin{minipage}{\textwidth}
    \centering
    \par \bigskip
    \includegraphics[width=0.7\textwidth]{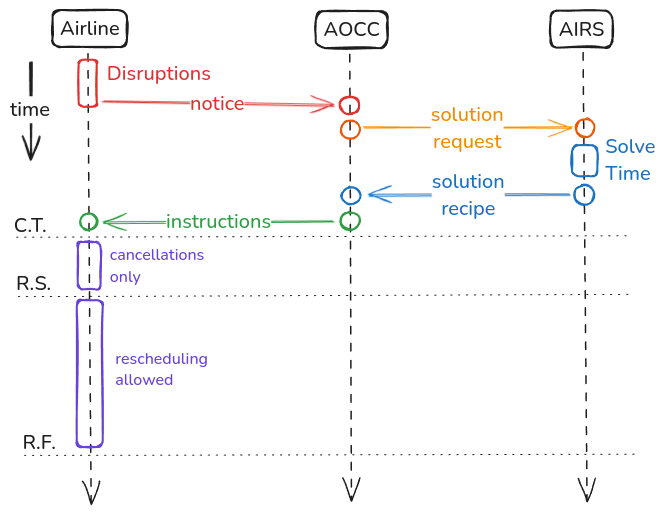}
    \caption{Approximate problem timeline, such that ``C.T.'' is Current Time, ``R.S.'' is Recovery Start, and ``R.F.'' is Recovery Finish. Not in scale.}
    \par \bigskip
    \label{fig:airs_timeline}
\end{minipage}
\end{figure}

\subsection{The AIRS method}

The AIRS method is composed of 2 main stages:
\begin{enumerate}
    \item The \textbf{AIRS.ACR} module finds a feasible schedule considering mainly Aircraft and Crew, but not considering Passenger Itineraries directly.
    \item The \textbf{AIRS.PaxR} module adjust the schedule generated by the previous module, keeping feasible the assignment of Aircraft \& Crew, and re-assigning passengers to the flights in the new itinerary.
\end{enumerate}

\begin{figure}[H]
\begin{minipage}{\textwidth}
    \centering
    \par \bigskip
    \includegraphics[width=\textwidth]{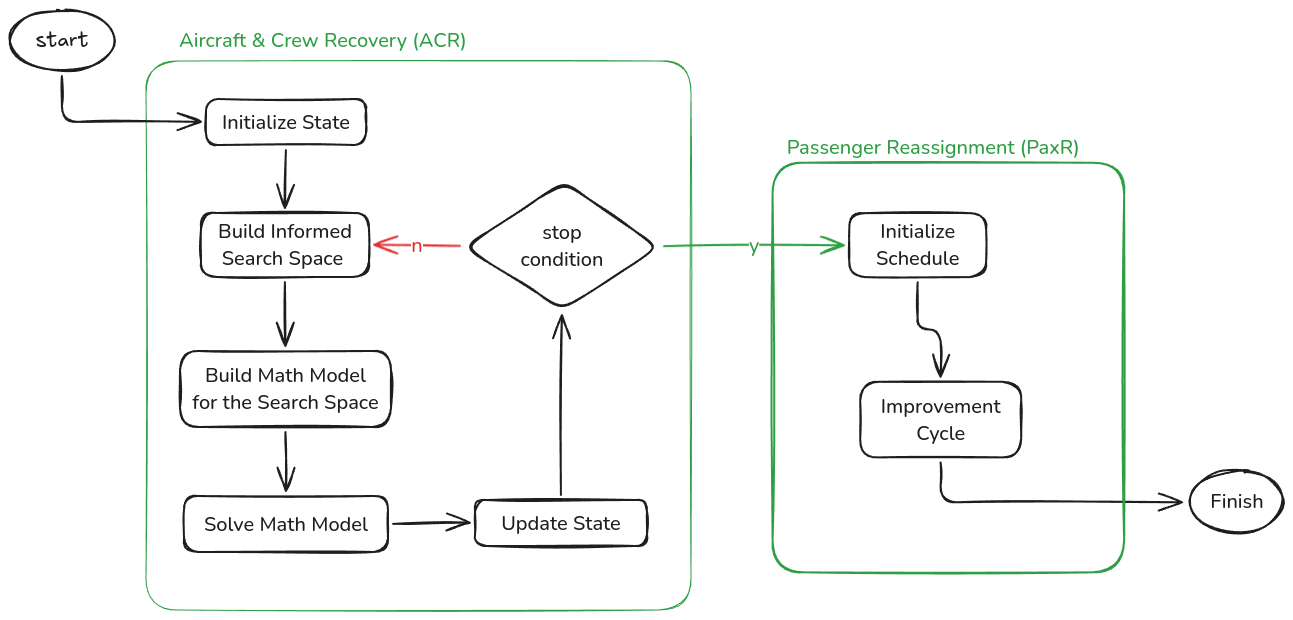}
    \caption{Flow diagram: top level view of solution steps in the AIRS solver.}
    \par \bigskip
    \label{fig:airs_top_level_diagram}
\end{minipage}
\end{figure}

Splitting the solution into a sequential two-part method reduces the time-to-solution and the required computational resources significantly (by reducing the combinatorial scaling in the number of variables and constraints). However, there are trade-offs involved. In particular, the division of Aircraft and Crew scheduling is usually the most critical.

Previous experiments showed that it is often very hard to reach feasible solutions in larger problem instances when the stages of aircraft rescheduling and crew rescheduling are split apart. This is not a cost issue, but rather a feasibility issue. A flight requires crew and aircraft to be available simultaneously and in accordance with many business rules. Computing both in the same mathematical model, in turn, tends to cause explosive growth in the constraint and variable counts, which means the model runs for much longer and uses much more memory. There is a trade-off between feasibility at larger problem sizes and computational complexity.

The approach adopted in \textbf{AIRS.ACR} focuses on containing the combinatorial growth of model size while still building an integrated model for aircraft and crew. The method can consider both resources at once, yielding a feasible schedule right in the first iteration, reducing cost over time. It has been shown to find feasible solutions within practical time limits even at larger problem sizes, and there remains scope for performance improvements. Empirically, this design achieves first-iteration feasibility with progressive cost reduction while maintaining tractable runtimes on large instances. The size of some tested problems is presented in Table \ref{tab:instances}.

On the other hand, splitting these critical resources from the passenger reassignment may yield solutions that disrupt passenger itineraries more than necessary. To handle this issue, the \textbf{AIRS.PaxR} module was designed to assign itineraries to flights while adjusting the results from \textbf{AIRS.ACR} to route more itineraries to their destinations, while maintaining schedule feasibility with respect to all constraints. Additional cost components are also considered at this stage.

\subsection{ACR: Aircraft \& Crew Recovery}

AIRS.ACR is composed of the following high-level modules:
\begin{itemize}[noitemsep]
    \item Data Preprocessing (Flights, Maintenance, Crews, Slots)
    \item Search Space Construction
    \item MILP Model Generation
    \item Solver Execution
    \item Output Interpretation and Visualization
\end{itemize}

\subsubsection{Mathematical Model Overview}

The core model is a MILP, in which each operational decision is represented by one or more binary variables. Higher-level structures are used to control the creation and retrieval of such variables. The main abstraction used for this representation is the \textbf{Choice}, of which there are two types:

\begin{itemize}[noitemsep]
    \item \textbf{OptionChoice}: represents state for flight or maintenance, such as Scheduled, Canceled, Succeeding Maintenance, Failing Maintenance;
    \item \textbf{SlotChoice}: represents airport slot usage with associated penalties for underutilization.
\end{itemize}

Each Choice is associated with a linear “cost expression”, calculated based on specific rules for each type of Choice. Using this abstraction, the optimization objective is defined as follows:
\[ \min \sum_{\text{choice}} \text{cost\_expr(choice)} \]

Costs in this objective will reflect:
\begin{itemize}[noitemsep]
    \item Delay minutes
    \item Cancelations
    \item Infeasible maintenances
    \item Slot underutilization when a penalty is specified
\end{itemize}

And the optimization will be subject to the following constraints:
\begin{itemize}

\item Flow balance across the Time Space Network:
\begin{align}
&\substack{
    \forall\ node\ \in\ \mathbf{each\_node}(tsn) \text{, where ``} tsn \text{'' is the Time Space Network:}
} \nonumber \\[5pt]
&\begin{cases}
\quad\quad \displaystyle\sum_{\mathclap{\substack{arc\ \in\\\mathbf{outputs}(node)}}}\mathbf{decision}(arc)\ =\
\sum_{\mathclap{\substack{arc\ \in\\\mathbf{inputs}(node)}}}\mathbf{decision}(arc) \quad\  \text{if multileg node}\\\\
\quad\quad \displaystyle\sum_{\mathclap{\substack{arc\ \in\\\mathbf{outputs}(node)}}}\mathbf{decision}(arc)\ \le\
\sum_{\mathclap{\substack{arc\ \in\\\mathbf{inputs}(node)}}}\mathbf{decision}(arc) \quad\ \text{otherwise}\\
\end{cases}
\end{align}
Multi-leg connections are handled in a different virtual position than usual connections, and the strict flow constraint guarantees that either all legs or none are performed.

\item Unique decision per flight/maintenance group (ensures that no flight is assigned to more than one combination of aircraft, crew, and time, or is cancelled):
\begin{align}
\substack{
\forall\ choice\_group\ \in\ tsn.option\_choices\_by\_entity \\ (\text{grouped by maint or flight})}\quad\quad
\sum_{\mathclap{\substack{choice\ \in\\choice\_group}}}\mathbf{decision\_expr}(choice)\ =\ 1
\end{align}

\item Slot capacity preservation (ensures that each slot is not used more than its capacity and makes sure the nonuse\_var of the \textbf{Slot Choice} has the correct value):
\begin{align}
&\substack{\forall\ choice\ \in\ tsn.slot\_choices \text{ and associated } slot:}\\\nonumber\quad\quad
&\mathbf{decision\_expr}(choice) + \mathbf{nonuse\_var}(choice) = \mathbf{capacity}(slot)
\end{align}

\item Crew flight time limits (ensures that each crew group isn't assigned to fly more than its set limit):
\begin{align}
&\scriptstyle{\forall\ crew\ \in\ \mathbf{crew\_groups}(tsn):}\displaystyle\nonumber\\\quad\quad
&\sum_{\mathclap{\substack{choice\ \in\\ \mathbf{flight\_choices}(crew)}}}\mathbf{decision}(choice)\mathbf{duration}(choice)\ \le\ 
\mathbf{flight\_time\_limit}(crew)
\end{align}

\end{itemize}

To take advantage of the constraints and costs presented above, the Time-Space Network \textbf{Nodes} and \textbf{Arcs} must be built following the rules detailed further.

\subsubsection{Time-Space Network Construction}

The \textbf{TSN} models resources (aircraft, crew) over time and space. \textbf{Nodes} represent possible states of a resource; \textbf{Arcs} represent state transitions, usually associated with actions over time such as flight legs or boarding a plane. There are several different rules on how to add \textbf{Arcs} to the networks, depending on the available \textbf{Choices}, as detailed in this section.

\textbf{\textit{Option Choices:}} Depending on the kind of \textbf{Option} being considered in the \textbf{Choice}, a different algorithm is needed for choosing \textbf{Arcs}:

\textbf{\textit{Scheduled Flight Option:}} When the \textbf{Option} being considered is to \textbf{Schedule} the flight, it's necessary to add \textbf{Arcs} that represent the flow between the airports for the aircraft and for the crew, as well as special \textbf{Arcs} to allow embarking the crew onto the aircraft in case they are not already embarked, and disembarking \textbf{Arcs} at the destination so that the crew has the chance to leave the flight. The embarking/disembarking \textbf{Arc} is not added in case the corresponding connection is a multi-leg transit. The minimum turnaround or transit is enforced by an extension on the tail of the aircraft's \textbf{Arc}, while the connection time for the crew is handled by the embarking \textbf{Arc} in case the crew needs to change aircraft.

The \textbf{Arc} associated with performing the flight on the aircraft's network will share the same decision variable as the one from the crew's network, and this variable is the only variable in the \textbf{Option Choice}'s decision expression. On the other hand, there are \textbf{Arcs} associated with embarking and disembarking crew, each with their own independent variables to account for crew movement that is not completely linked to the execution of the flight. Thus, if both \textbf{Arcs} can carry sufficient flow, the \textbf{Option} is feasible.

\begin{figure}[H]
\begin{minipage}{\textwidth}
    \centering
    \par \bigskip
    \includegraphics[width=0.8\textwidth]{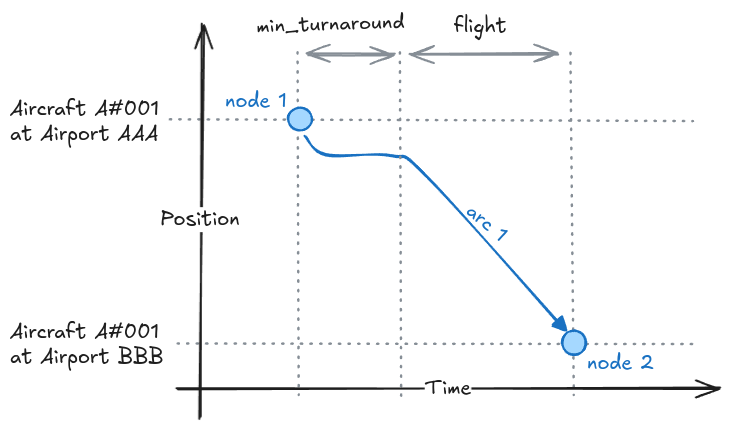}
    \caption{Required \textbf{Arcs} for representing the \textbf{Option} of scheduling a flight in the \textbf{aircraft} network.}
    \par \bigskip
    \label{fig:sched_flight_option_aircraft_arcs}
\end{minipage}
\end{figure}

\begin{figure}[H]
\begin{minipage}{\textwidth}
    \centering
    \par \bigskip
    \includegraphics[width=0.8\textwidth]{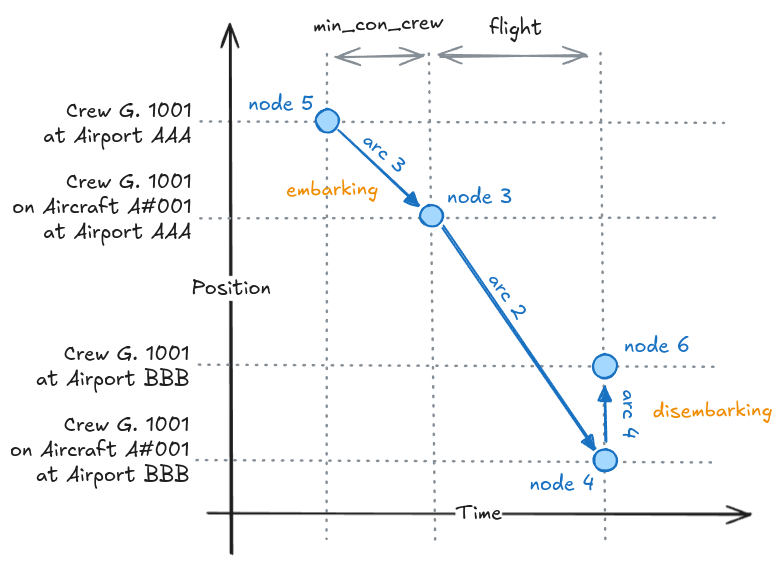}
    \caption{Required \textbf{Arcs} for representing the \textbf{Option} of scheduling a flight in the \textbf{crew} network.}
    \par \bigskip
    \label{fig:sched_flight_option_crew_arcs}
\end{minipage}
\end{figure}

A simple example of the required \textbf{Arcs} is shown in Figures \ref{fig:sched_flight_option_aircraft_arcs} and \ref{fig:sched_flight_option_crew_arcs}. Note that \textbf{Arc 3} and \textbf{Node 5} will not be created for this flight if it is part of a multi-leg and not the first leg, while \textbf{Arc 4} and \textbf{Node 6} will not be created for this flight if it is part of a multi-leg which does not conclude at that point. In the first case, the minimum transit would be used instead of the minimum turnaround.

\begin{figure}[H]
\begin{minipage}{\textwidth}
    \centering
    \par \bigskip
    \includegraphics[width=\textwidth]{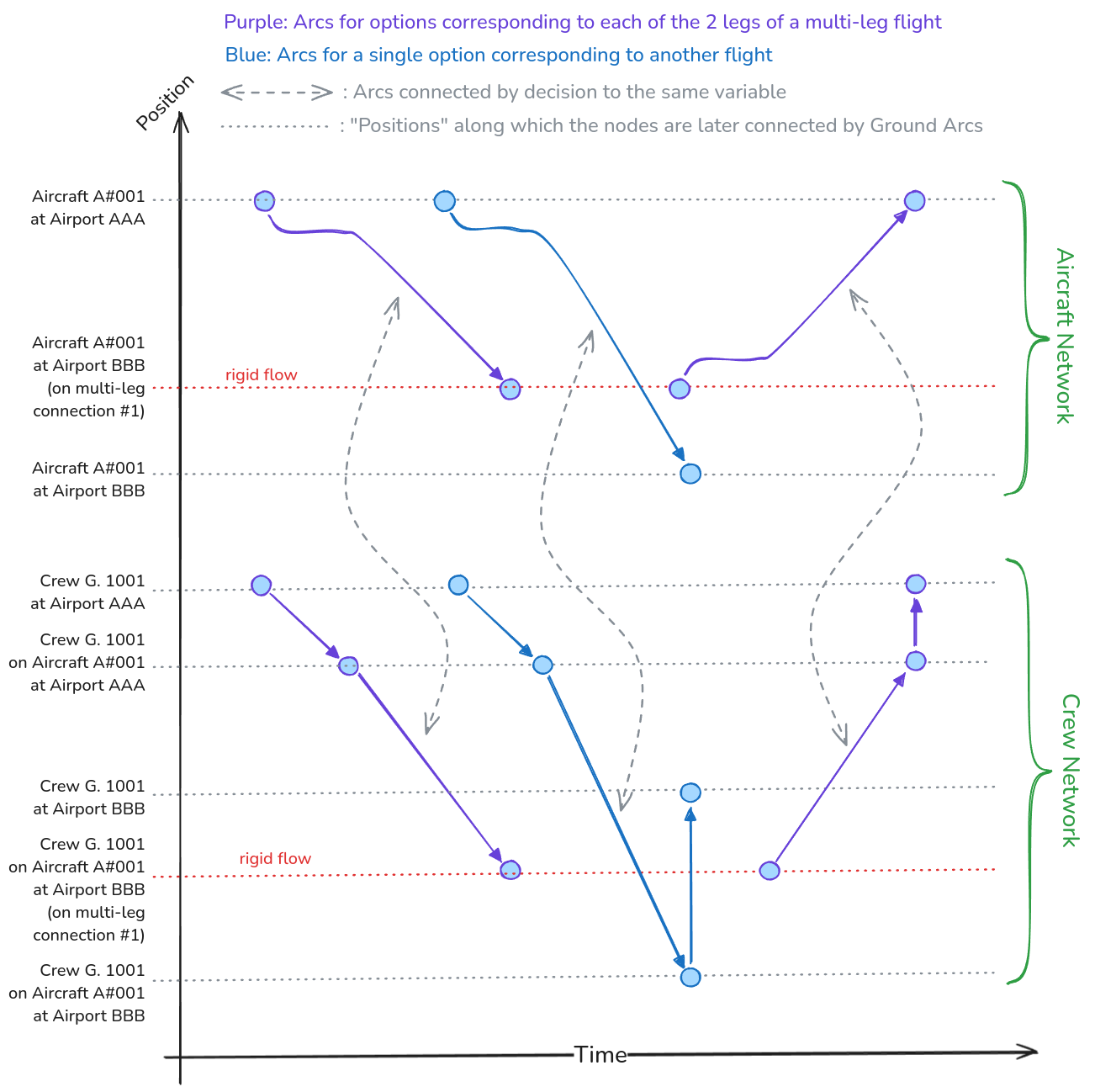}
    \caption{Example scheme of \textbf{Arcs} for one 2-leg flight and one single-leg flight.}
    \par \bigskip
    \label{fig:multileg_example}
\end{minipage}
\end{figure}

To ensure that, for example, an aircraft performs only the first leg of a multi-leg flight and then proceeds to operate a completely different flight, there are special positions to describe being at a particular airport during the transit between two specific legs of a multi-leg flight. We refer to these virtually separate positions as \textit{sub-threads}''. These are treated as completely separate places when adding the \textbf{Ground Arcs} later. At those positions, the flow balance constraint is also more rigid'' ($=$ instead of $\le$), which means that every unit of flow that goes in must go out. This guarantees that either all the legs of a multi-leg flight are canceled, or they are all assigned to the same aircraft and crew. Figure \ref{fig:multileg_example} shows an example with a two-leg flight and a single-leg flight.

% FILIP: perhaps we could differentiate (e.g. with node colors?) arcs related to crew and the aircraft itself? Unless we intentionally want to communicate, that all of these are "arcs" and we do not want to differentiate between these at that point, but in such case it could be worth pointing that out to the Reader. 
%-----> Does this look good enough?

Also, note how the disembarking \textbf{Arcs} are vertical in Figures \ref{fig:sched_flight_option_aircraft_arcs} and \ref{fig:multileg_example}. In practice, they connect at the same time on the main thread. Consequently, this model requires the crew's minimum connection time to be greater than zero.

\begin{figure}[H]
\begin{minipage}{\textwidth}
    \centering
    \par \bigskip
    \includegraphics[width=0.7\textwidth]{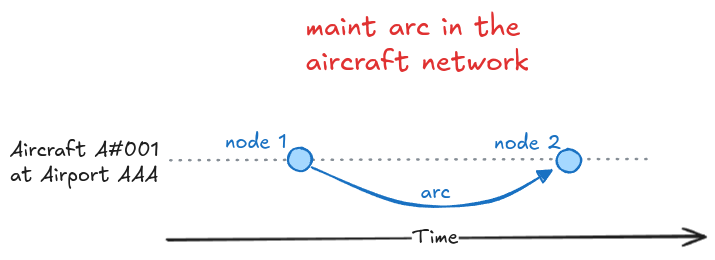}
    \caption{\textbf{Arc} representation of the \textbf{Option} of successfully performing a maintenance at a given time and airport.}
    \par \bigskip
    \label{fig:succeeding_maint_arc}
\end{minipage}
\end{figure}

\textbf{\textit{Succeeding Maintenance:}} The \textbf{Succeeding} state holds information about the location and time for this maintenance. Within a single iteration of the ACR method, each maintenance will be associated with many \textbf{Options} with a different \textbf{Succeeding} states, each with distinct times and locations, representing different ways for that maintenance to be performed successfully. In the previous stage, the search space construction, the \textbf{Maintenance Options} are chosen based on the set of \textbf{Flight Options} under consideration. It is possible to represent each of these individual \textbf{Options} with a single \textbf{Arc} in the aircraft network, as shown in Figure \ref{fig:succeeding_maint_arc}.

\begin{figure}[H]
\begin{minipage}{\textwidth}
    \centering
    \par \bigskip
    \includegraphics[width=0.9\textwidth]{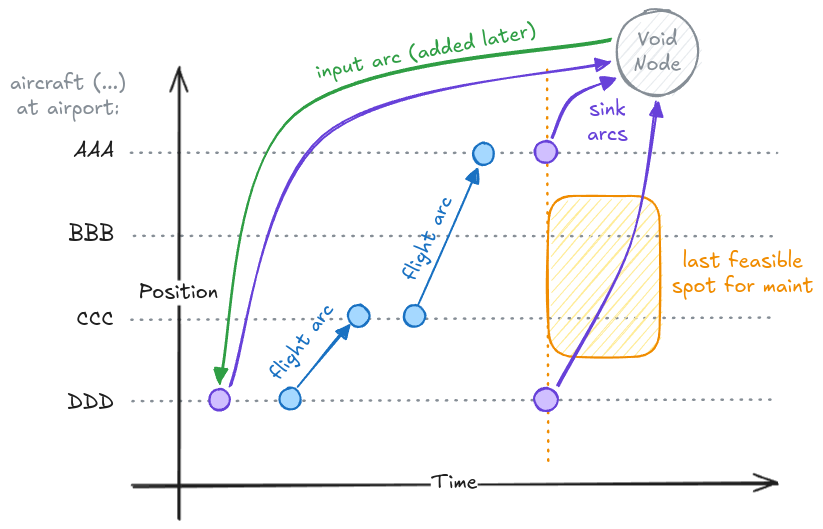}
    \caption{Required sink \textbf{Arcs} required to represent the \textbf{Option} of failing to perform a maintenance in a given aircraft, excluding the possibility of performing further flights with this vehicle.}
    \par \bigskip
    \label{fig:maint_fail_arcs}
\end{minipage}
\end{figure}

\textbf{\textit{Failing Maintenance:}} While omission of required maintenance is undesirable and would ideally be prohibited, strict enforcement can render the problem infeasible under severe disruption conditions; therefore, a soft-constraint mechanism is employed. Specifically, we assign a high penalty to the \textbf{Option} of failing maintenance and represent it via a set of \textbf{Sink Arcs}, as illustrated in Figure \ref{fig:maint_fail_arcs}. These sink \textbf{Arcs} impose no flow-conservation constraint at the terminal \textbf{Node} by terminating at a \textbf{Void Node}; this prevents further assignments while recording the violation as a penalized choice. The \textbf{Void Node} serves as an origin and/or destination wherever unconstrained flow is required and is likewise used for the next two categories of \textbf{Arcs}.

\textbf{\textit{Canceling Flight:}} This is the simplest case among the \textbf{Option Choices}, as it does not require any flow restriction. Thus, no \textbf{Arc} is required in any network, and free variables may be used instead. For debugging and data analysis, it may be useful to generate \textbf{Arcs} that start and finish at the \textbf{Void Node}, so as to retain those \textbf{Options} represented in the network data.

\begin{figure}[H]
\begin{minipage}{\textwidth}
    \centering
    \par \bigskip
    \includegraphics[width=0.9\textwidth]{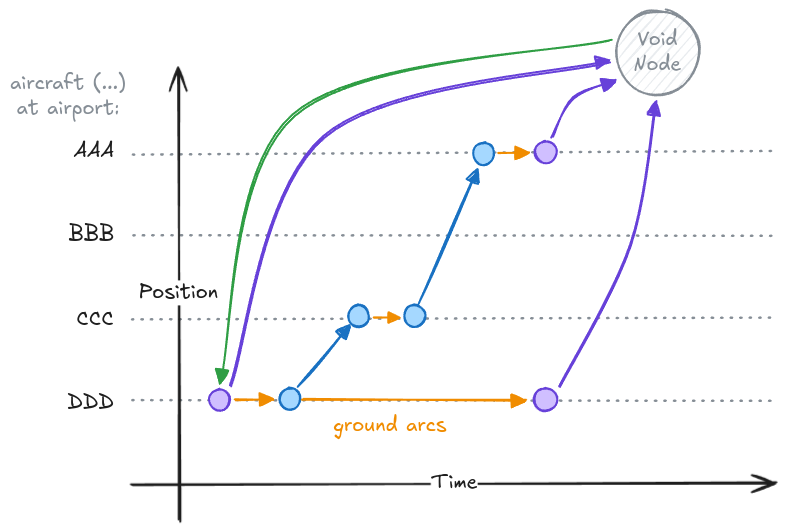}
    \caption{Generated \textbf{Ground Arcs} (in orange) and \textbf{Input Arc} (in green) for the network shown in figure \ref{fig:maint_fail_arcs}.}
    \par \bigskip
    \label{fig:input_and_ground}
\end{minipage}
\end{figure}

%FILIP: I've noticed that some parts
%-----> ?
\textbf{\textit{Input Arcs:}} We represent the initial position for an Aircraft or Crew with an \textbf{Arc} that starts at the \textbf{Void Node}, ends at the position/time at which the input is given, and has a fixed decision (flow) of 1. Figure \ref{fig:input_and_ground} presents an example input \textbf{Arc} (in green). These arcs work exactly the same way for both Aircraft and Crew. If the resource is reported as present at that airport but no previous history of flights is given, it will be assumed to have been there for an arbitrarily long time before \textbf{Current Time}. Otherwise it will be placed at the finishing position/time of its last flight starting before \textbf{Current Time}.

\textbf{\textit{Ground Arcs:}} These are added after all the others to connect the timelines in each position'' in the network. For example, consider the network presented earlier for the maintenance failure case. Its generated \textbf{Ground Arcs} are shown in Figure \ref{fig:input_and_ground}. To generate such \textbf{Arcs}, the program groups the \textbf{Nodes} by their positions'', sorts the groups by time, and connects each pair of consecutive \textbf{Nodes} with a new \textbf{Arc} that has an independent variable.

\subsection{PaxR: Passenger Reassignment and Schedule Improvement}

The \textbf{AIRS.PaxR} module is responsible for reassigning passengers to flights. After finding a feasible flight schedule with the ACR module, the PaxR module uses a custom greedy algorithm to fix infeasible itineraries, thereby modifying the schedule to reduce costs associated with itinerary disruptions without violating already satisfied constraints.

The schedule provided by the ACR module may render some itineraries infeasible, but, in general, this is not expected to occur for the majority of itineraries. Other existing methods apply greedy algorithms \citet{bisaillon2010}, evaluating itineraries in priority order, keeping as many as possible unchanged and reassigning when needed to other routes. Algorithms \ref{alg:assign_itineraries}, \ref{alg:assign_fixed_itineraries}, \ref{alg:assign_feasible_itineraries}, and \ref{alg:assign_infeasible_itineraries} present pseudocode that provides a high-level view of how the algorithm operates. Algorithm \ref{alg:assign_itineraries} displays the order in which the main operations are performed, and the others give additional detail on the operation of those procedures. The main procedure initializes flight capacities and assigns passengers in three passes--first preserving fixed portions of itineraries, then filling feasible original paths, and finally routing remaining demand along the earliest viable alternatives--updating capacities after each step. The subroutines prioritize earliest arrival and minimal downgrades while respecting remaining capacity, thereby reducing cancellations and preserving service where possible.

\begin{algorithm}[H]
\caption{Assign Itineraries (Main Procedure)}
\label{alg:assign_itineraries}
\begin{algorithmic}[1]
\Procedure{AssignItineraries}{}
    \State initialize $Flights$ with max capacity
    \State \text{AssignFixedItineraries}(\textit{flights}, \textit{itineraries}, ...)
    \State \text{AssignFeasibleItineraries}(\textit{flights}, \textit{itineraries}, ...)
    \State \text{AssignInfeasibleItineraries}(\textit{flights}, \textit{itineraries}, ...)
\EndProcedure
\end{algorithmic}
\end{algorithm}

\begin{algorithm}[H]
\caption{Assign Fixed Itineraries}
\label{alg:assign_fixed_itineraries}
\begin{algorithmic}[1]
\Procedure{AssignFixedItineraries}{\textit{flights}, \textit{itineraries}, ...}
    \ForAll{\textit{itinerary} starting before RP}
        \State \(\textit{path} \gets\) original pre-RP flights assigned up to the last feasible (no downgrade)
        \State \(\textit{capacity} \gets\) minimum capacity of flights in \textit{path} (original cabin classes)
        \If{\(\textit{path}\) not empty and \textit{capacity} non-zero}
            \State assign \textit{itinerary} to all flights in \textit{path} (original cabin classes)
            \State subtract \textit{capacity} from all flights in \textit{path} (original cabin classes)
        \EndIf
    \EndFor
    \Statex
    \State \(\textit{partiallyAssigned} \gets\) itineraries assigned to parts of their original trips on the loop above
    \State Sort \(\textit{partiallyAssigned}\) by estimated remaining trip time (ascending)
    \Statex
    \ForAll{\textit{itinerary} in \textit{partiallyAssigned}}
        \State \(\textit{remainingPath} \gets\) rest of the original flights up to the last feasible (no downgrade)
        \State \(\textit{capacity} \gets\) minimum capacity of flights in \textit{remainingPath} (original cabin classes)
        \If{\textit{remainingPath} not empty and \textit{capacity} non-zero}
            \State assign \textit{itinerary} to all flights in \textit{remainingPath}
            \State subtract \textit{capacity} from all flights in \textit{remainingPath}
        \EndIf
        \Statex
        \While{\textit{itinerary} has unassigned passengers}
            \State \(\textit{bestPath} \gets\) path of earliest arrival to destination (least downgrade possible)
            \State \(\textit{capacity} \gets\) minimum capacity of flight assignments in \textit{bestPath}
            \If{\(\textit{bestPath}\) is not found}
                \State \textbf{break}
            \EndIf
            \State apply all assignments in \textit{bestPath} to \textit{itinerary}
            \State subtract \textit{capacity} from all flights in \textit{bestPath} accordingly to the cabin class
        \EndWhile
    \EndFor
\EndProcedure
\end{algorithmic}
\end{algorithm}

\begin{algorithm}[H]
\caption{Assign Feasible Itineraries}
\label{alg:assign_feasible_itineraries}
\begin{algorithmic}[1]
\Procedure{AssignFeasibleItineraries}{\textit{flights}, \textit{itineraries}, ...}
    \State \(\textit{itineraries} \gets\) itineraries not treated by AssignFixedItineraries
    \ForAll{\textit{itinerary} in decreasing order of estimated cancellation costs}
        \State Determine how many can be scheduled on original flights without downgrading
        \If{any can be scheduled}
            \State Assign them and update \(\textit{flight capacities}\)
        \EndIf
    \EndFor
\EndProcedure
\end{algorithmic}
\end{algorithm}

\begin{algorithm}[H]
\caption{Assign Infeasible Itineraries}
\label{alg:assign_infeasible_itineraries}
\begin{algorithmic}[1]
\Procedure{AssignInfeasibleItineraries}{\textit{flights}, \textit{itineraries}, ...}
    \State \(\textit{itineraries} \gets\) itineraries not treated by Assign(Fixed/Feasible)Itineraries
    \ForAll{\textit{itinerary} in decreasing order of estimated cancellation costs}
        \State \(\textit{best\_path} \gets\) find earliest arrival path with remaining capacity
        \If{\(\textit{best\_path}\) is found}
            \State Assign as many passengers as possible to \(\textit{best\_path}\), minimizing downgrades
            \State Update \(\textit{flight capacities}\)
        \EndIf
    \EndFor
\EndProcedure
\end{algorithmic}
\end{algorithm}

At the same time, some evolutionary method may be applied to change the schedule progressively in order to allow more passengers to complete their trips and further reduce costs.
As the greedy algorithm is very fast to execute, thousands of possible schedules may be evaluated in a few minutes.
PaxR applies a custom genetic algorithm %NOTE crossover operator not implemented yet, but may be included soon, tests on the way
that evaluates individuals in each generation in parallel.
Given a baseline schedule taken from the solution obtained by the ACR module, the internal representation of each individual to be evolved is a pair of two elements:

\begin{enumerate}
    \item \textbf{ScheduleCandidate}: a data structure that contains a set of modifications to the baseline schedule
    \item \textbf{PaxAssignmentCandidate}: a different data structure that contains passenger assignment compatible withe baseline schedule after the modifications specified in \textbf{ScheduleCandidate} are applied.
\end{enumerate}

The only references shared among different individuals are the associated problem and baseline schedule, stored in the \textit{PaxR} structure, which is not mutated after construction. Effectively, this means we can safely work with these individuals in a multithreaded environment with shared memory without race conditions.

Due to the heavy use of structs, careful memory management is important for this algorithm. Julia's built-in garbage collector caused crashes when it was not explicitly called in the evolution loop, due to not correctly handling the cleanup of the large number of leftover structs removed from the population during the iterations.

Other than that, the evolutionary algorithm applied was a very simple non-crossover genetic algorithm:

\begin{algorithm}[H]
\caption{PaxR (non-crossover GA)}
\label{alg:evo_sched_opt}
\begin{algorithmic}[1]
\State $\mathcal{P} \gets \{\text{baseline schedule with passenger assignment}\}$
\State Generate $49$ mutated schedules; \quad $\mathcal{P} \gets \mathcal{P} \cup \{\text{mutations}\}$
\ForAll{$x \in \mathcal{P}$}
  \State $f(x) \gets \text{Cost}(x)$
\EndFor
\While{\text{termination criterion not met}}
  \State Select $30$ schedules from $\mathcal{P}$ with probability biased toward lower cost
  \ForAll{$p$ in selected set} \Comment{can execute in parallel}
    \State $c \gets \textsc{Mutate}(p)$ \Comment{respect all constraints}
    \State $\textsc{AssignPassengers}(c)$
    \State $f(c) \gets \text{Cost}(c)$
    \State $\mathcal{C} \gets \mathcal{C} \cup \{c\}$
  \EndFor
  \State $\mathcal{P} \gets \text{Best}_{50}\big(\mathcal{P} \cup \mathcal{C}\big)$ \Comment{ascending cost}
\EndWhile
\end{algorithmic}
\end{algorithm}

\begin{enumerate}[nosep]
    \item starts with a population of 1 (no change to the baseline, and the associated passenger assignment);
    \item creates 49 mutated schedules to compose an initial population of 50 individuals;
    \item calculates the estimated costs of every individual;
    \item finally, on each evolution iteration: \begin{enumerate}[nosep]
        \item choose 30 schedules from the current population from a distribution that favors lower costs;
        \item the following is done in multithreading for each of the 30 schedules chosen: \begin{enumerate}[nosep]
            \item create a mutated version of the schedule, keeping all constraints respected;
            \item create the associated passenger assignment;
            \item calculate the associated cost;
        \end{enumerate}
        \item next population will be the 50 individuals with lower costs among all the 80 composed by the previous population and the new mutations.
    \end{enumerate}
\end{enumerate}

The size of the population, the allowed optimization time, as well as the choice of evolutionary method can be changed in order to better utilize available hardware and to better suit user requirements.

\section{Business Constraints and Data Integration}

This family of constraints consists mainly of the following:

\begin{itemize}[noitemsep]
    \item Aircraft rotation consistency
    \item Crew legality and duty limits
    \item Airport slot capacity and overlapping
    \item Maintenance location/time constraints
\end{itemize}
These constraints are injected during search space and TSN construction phases.

\section{Solver Strategy and Implementation}

The solver builds the TSN, encodes decisions as symbolic variables, and uses MILP solvers (e.g., Gurobi or CBC). Key strategies:

\begin{itemize}[noitemsep]
    \item Warm start from current schedule
    \item Penalty tuning for preference control
    \item Layered constraint enforcement
\end{itemize}

\section{An Example Problem on AIRS.ACR}

\begin{figure}[H]
\begin{minipage}{\textwidth}
    \centering
    \par \bigskip
    \includegraphics[width=0.9\textwidth]{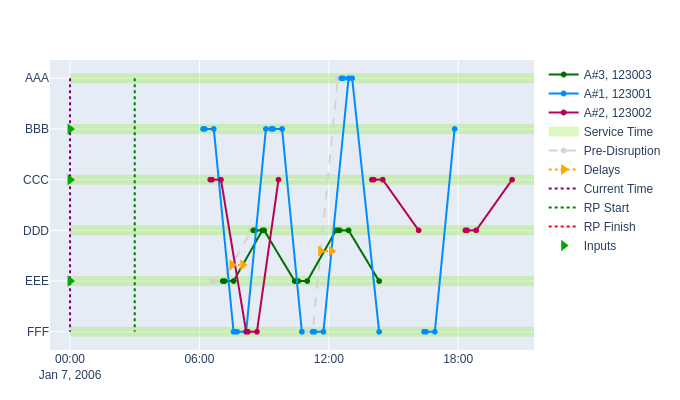}
    \caption{\small Example aircraft \& crew rescheduling problem, split by the original aircraft assigned to each flight (A\#1, A\#2 and A\#3), and by the original crew groups (123001, 123002 and 123003). Recovery Finish is ahead of all the events in this problem and is not shown in this plot.}
    \par \bigskip
    \label{fig:example_problem}
\end{minipage}
\end{figure}

In this section, a small problem is visualized to ilustrate the sequence of steps executed by the ACR module. This example problem consists of a short schedule, illustrated in Figure \ref{fig:example_problem} with:

\begin{itemize}[noitemsep]
    \item 7 airports
    \item 3 aircraft
    \item 5 crew groups
    \item 14 flights
    \item 2 initial flight delays
    \item everything happening after Recovery Start, for simplicity
\end{itemize}

%FILIP: Again, we need a bit more control over the figure placement.
%-----> TODO go over the entire layout after the contents are all fixed

\begin{figure}[H]
\begin{minipage}{\textwidth}
    \centering
    \par \bigskip
    \includegraphics[width=0.9\textwidth]{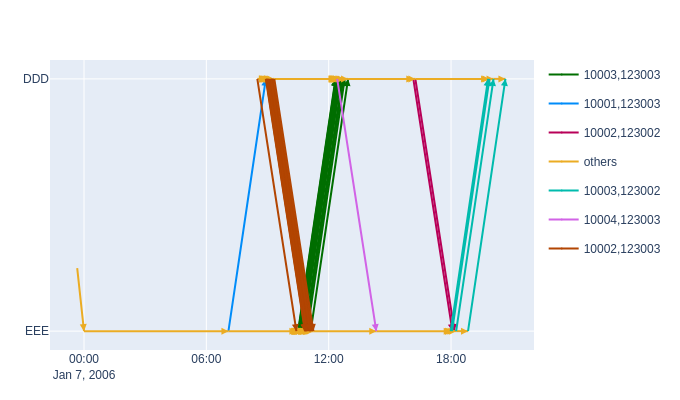}
    \caption{\small Time-Space Network segment generated for aircraft A\#3 in the first iteration of the ACR method. The related problem is the one ilustrated in Figure \ref{fig:example_problem}. In the legend, the first number is the identifier of the flight leg, while the second number is the identifier of a crew group. Each arrow represents an \textbf{Option} for performing a given flight with a given crew group on aircraft A\#3. The ``others'' label refers to \textbf{Arcs} that do not represent \textbf{Options} directly linked to flights, such as \textbf{Ground Arcs} and the \textbf{Input Arc} in the bottom-left. Other aircraft have their own parts in the overall Time-Space Network.}
    \par \bigskip
    \label{fig:tsn_plot_aircraft}
\end{minipage}
\end{figure}

\begin{figure}[H]
\begin{minipage}{\textwidth}
    \centering
    \par \bigskip
    \includegraphics[width=0.9\textwidth]{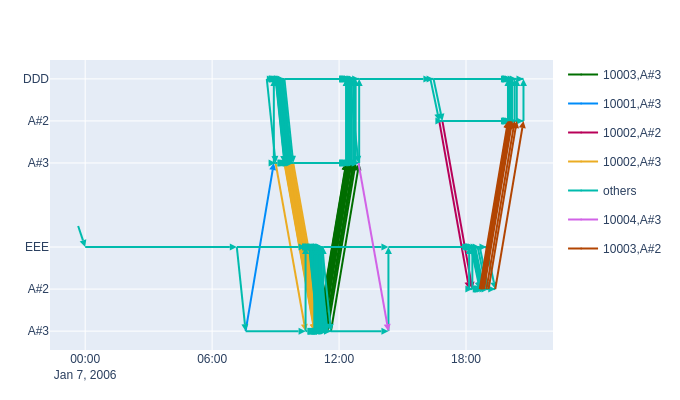}
    \caption{\small The Time-Space Network part generated for crew 123003 in the first iteration of the ACR method. The related problem is the one illustrated in Figure \ref{fig:example_problem}. In the legend, the first number is the identifier of the flight leg, while the second number is the identifier of an aircraft. Each arrow represents one \textbf{Option} for performing a given flight with a given aircraft and crew 123003. The ``others'' label refers to \textbf{Arcs} that do not represent \textbf{Options} directly linked to flights, such as ground \textbf{Arcs}, the input \textbf{Arc} on the far left, and the embark/disembark \textbf{Arcs} under each airport. It is worth noting the inclusion of vertical positions for the aircraft under each airport, which allows skipping connection time when staying on the same aircraft. Other crew groups will have their own parts in the overall Time-Space Network.}

    \par \bigskip
    \label{fig:tsn_plot_crew}
\end{minipage}
\end{figure}

The method applied in the ACR module is iterative, and at each iteration, a Time-Space Network is built. The network is divided into subnetworks associated with each aircraft and each crew, linked to the set of associated \textbf{Options} in the search space. Figures \ref{fig:tsn_plot_aircraft} and \ref{fig:tsn_plot_crew} show the time-space generated for, respectively, a given aircraft and a given crew group, specifically in the first iteration.

To contain the rapid growth in the number of variables, the method avoids enumerating all flight–crew–time combinations and instead activates \textbf{Options}, represented as arrows, concentrated around observed irregularities. In the figures, the disruptions depicted are those detectable from the baseline schedule; the first iteration therefore targets these visible events. In subsequent iterations, solver feedback reveals additional disruptions -- such as large delays, flight cancellations, and maintenance infeasibilities -- these events guide the injection of new \textbf{Options}, focusing computation on decisions that drive feasibility and cost.

\begin{figure}[H]
\begin{minipage}{\textwidth}
    \centering
    \par \bigskip
    \includegraphics[width=0.9\textwidth]{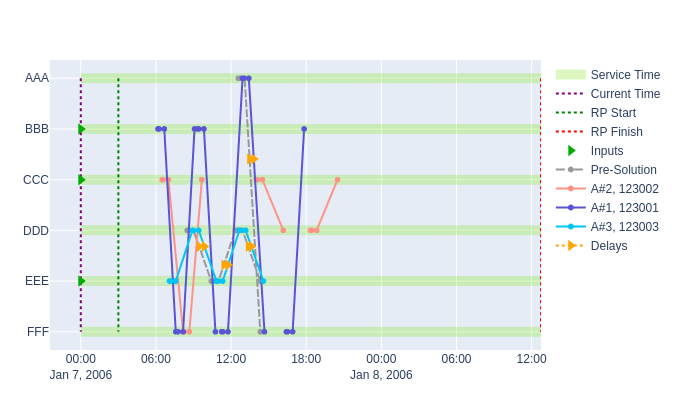}
    \caption{\small Solution achieved by the ACR method for the problem illustrated in Figure \ref{fig:example_problem}. No cancellations or swaps are required, although alternative crews are available. Some delays are needed to account for those in the original problem. This solution is reached in the second iteration because of the chain of delays on aircraft A\#3, the initial iteration usually cancels the last flight. This cancellation is then solved in the second iteration.}
    \par \bigskip
    \label{fig:solution_plot}
\end{minipage}
\end{figure}

This small problem can be solved to a satisfactory level in about 2 iterations, adding as few as 100 change \textbf{Options} to the system, in around 10 seconds on our test hardware, which is specified in Section \ref{sec:experiments}.
Solving larger problems with more disruptions requires a larger number of change \textbf{Options} per iteration, as well as more iterations.
Figure \ref{fig:solution_plot} shows the solution reached for this specific problem, which requires a small number of delays.

\section{Experimental Results}\label{sec:experiments}

This section presents statistics on the execution of AIRS on 5 distinct test instances sized similarly to real-world problems. The recovery time windows are approximately one day. Detailed information on the nature of each instance is presented in Table \ref{tab:instances}. Tables \ref{tab:acr_bench}, \ref{tab:paxr_bench}, and \ref{tab:global_bench} contain information on time-to-solution for several different parts of the solver. The tables do not include statistics for feasibility, as the solver achieved feasible solutions for all the tested instances.

Note that AIRS is highly configurable in terms of cost and required TTS. This benchmark was executed using the current default parameters, targeting an execution time of around 10 minutes, as reflected in the third column of Table \ref{tab:global_bench}. The stopping conditions for ACR and PaxR adhere to this time target, seeking the best solution reachable within the available time. Therefore, the sub-step times are generally more informative.

It is also worth noting that the last column in Table \ref{tab:paxr_bench} reports the number of generations for the evolutionary algorithm applied to the schedule. This evolutionary algorithm uses, by default, a population size of 50 schedules per generation, with 30 mutations per iteration. These settings are adjustable, and alternative heuristic strategies can also be employed (e.g., simulated annealing and other derivative-free methods). In practice, the population size and mutation rate can be tuned to balance runtime and solution quality.

The computational experiments were performed on a system with the specifications detailed below:

\begin{table}[H]
  \centering
  \caption{System configuration used in computational experiments}
  \label{tab:systemspecs}
  \begin{tabular}{@{}ll@{}}
    \toprule
    Component     & Specification \\
    \midrule
    Processor     & Intel Core i9-9960X \\
    System memory & \SI{125.48}{GiB} \\
    Operating system & Ubuntu 22.04.5 LTS \\
    Software      & Julia 1.11.5 \\
    Parallelism   & Up to 12 threads \\
    GPU           & Not used (no acceleration required) \\
    \bottomrule
  \end{tabular}
\end{table}

% \begin{itemize} \label{systemspecs}
%     \item \textbf{Processor:} Intel Core i9-9960X
%     \item \textbf{System Memory:} 125.48 GiB
%     \item \textbf{OS:} Ubuntu 22.04.5 LTS
%     \item \textbf{Software:} Julia Version 1.11.5
%     \item \textbf{Parallelism:} A maximum of 12 threads were utilized for the computations.
%     \item \textbf{GPU:} Not utilized, as the current version of the algorithm does not require GPU acceleration.
% \end{itemize}

\begin{table}[H]
\centering
\caption{Entity counts for Instances A01-A07}
\label{tab:instances}
\sisetup{round-mode=places, round-precision=0, group-digits=false} % Setup for integer counts
\begin{tabular}{l
                S[table-format=2.0] % Airports
                S[table-format=2.0] % Slotted Airports
                S[table-format=2.0] % Aircraft
                S[table-format=3.0] % Crew Groups
                S[table-format=3.0] % Flights
                S[table-format=5.0] % Passenger Tickets
                S[table-format=1.0] % Multileg Connections
                S[table-format=3.0] % Flight Disruptions
                S[table-format=1.0] % Maintenances
                S[table-format=4.0]} % Airport Slots
\toprule
\multirow{1}{*}{Instance} &
\multicolumn{1}{c}{\rotatebox{90}{Airports}} &
\multicolumn{1}{c}{\rotatebox{90}{Slotted Airp.}} &
\multicolumn{1}{c}{\rotatebox{90}{Aircraft}} &
\multicolumn{1}{c}{\rotatebox{90}{Crew Groups}} &
\multicolumn{1}{c}{\rotatebox{90}{Flights}} &
\multicolumn{1}{c}{\rotatebox{90}{Pass. Tickets}} &
\multicolumn{1}{c}{\rotatebox{90}{Multileg Conn.}} &
\multicolumn{1}{c}{\rotatebox{90}{Flight Disrup.}} &
\multicolumn{1}{c}{\rotatebox{90}{Maint.}} &
\multicolumn{1}{c}{\rotatebox{90}{Airp. Slots}}
\\
\midrule
A01 & 35 & 25 & 85 & 162 & 608 & 43964 & 4 & 63  & 3 & 1478 \\
A02 & 35 & 29 & 85 & 162 & 608 & 43965 & 4 & 107 & 3 & 1478 \\
A04 & 35 & 32 & 85 & 162 & 608 & 43963 & 4 & 41  & 3 & 1478 \\
A06 & 35 & 29 & 85 & 162 & 608 & 54687 & 4 & 63  & 3 & 1478 \\
A07 & 35 & 30 & 85 & 162 & 608 & 54687 & 4 & 107 & 3 & 1478 \\
\bottomrule
\end{tabular}
\end{table}

\begin{table}[H]
\centering
\caption{ACR Performance Averages (Sub-step times are per ACR iteration)}
\label{tab:acr_bench}
\sisetup{round-mode=places, group-digits=false} % General siunitx settings for these tables
\begin{tabular}{lSSSSS}
\toprule
Avg. TTS$\rightarrow$ & {Prox.} & {S.S. Gen.} & {TSN Const.} & {TSN Optim.} & {Entire Iteration} \\
\LDtextarrow Instance & {(s)}   & {(s)}       & {(ms)}       & {(s)}      & {(s)} \\
\midrule
A01 & 6.2    & 22.5   & 466.0 & 8.6   & 37.8 \\
A02 & 6.2    & 22.3   & 428.6 & 5.3   & 34.2 \\
A04 & 17.9   & 24.0   & 442.3 & 11.9  & 54.2 \\
A06 & 16.5   & 23.3   & 430.9 & 8.6   & 48.8 \\
A07 & 16.6   & 23.1   & 431.4 & 5.0   & 45.1 \\
\bottomrule
\end{tabular}
\end{table}

\begin{table}[H]
\centering
\caption{PAXR Performance Averages (Times are per run)}
\label{tab:paxr_bench}
\sisetup{round-mode=places, group-digits=false}
\begin{tabular}{lSS}
\toprule
Avgs$\rightarrow$     & {Sched. Improv.} & {Evolution} \\
\LDtextarrow Instance & {TTS (s)}        & {Generations} \\
\midrule
A01 & 240.1 & 1156.6 \\
A02 & 240.1 & 1146.2 \\
A04 & 240.1 & 816.7 \\
A06 & 240.2 & 808.5 \\
A07 & 240.2 & 733.3 \\
\bottomrule
\end{tabular}
\end{table}

\begin{table}[H]
\centering
\caption{Benchmark Summary (Averages are per run)}
\label{tab:global_bench}
\sisetup{round-mode=places, group-digits=false}
\begin{tabular}{lSSSrr}
\toprule
         & Avgs$\rightarrow$ & {Full TTS} & {ACR} & {Initial} & {Final} \\
Instance & Runs              & {(s)}  & {Iters/Run} & {PaxR Cost}   & {PaxR Cost} \\
\midrule
A01 & 10 & 555.6 & 8.3 & 1242860 & 1095620 \\
A02 & 10 & 551.9 & 9.1 & 11247100 & 8556800 \\
A04 & 9  & 559.7 & 5.9 & 6330020 & 5561000 \\
A06 & 10 & 567.7 & 6.7 & 5942130 & 3652770 \\
A07 & 10 & 556.3 & 7.0 & 22047600 & 10678000 \\
\bottomrule
\end{tabular}
\end{table}

\section{Conclusion}

This paper presents AIRS.ACR, a novel solver-based engine for integrated disruption recovery in airline operations. By leveraging a Time-Space Network formulation and mixed-integer linear programming, AIRS.ACR enables simultaneous optimization of aircraft and crew schedules, addressing a long-standing challenge in airline operations control. Unlike traditional sequential methods, which often yield suboptimal or infeasible solutions due to resource decoupling, AIRS.ACR captures the complex interdependencies between aircraft and crew, producing more realistic and operationally viable recovery plans.

The system demonstrates strong scalability and computational efficiency, solving real-world-sized instances within practical timeframes. Its modular architecture -- separating aircraft and crew recovery (ACR) from passenger reassignment (PaxR) -- balances feasibility and performance, while allowing further refinement through evolutionary algorithms.

Experimental results confirm the system's ability to generate high-quality solutions under diverse disruption scenarios, with significant reductions in total disruption costs. Moreover, the solver's flexibility to incorporate business constraints (e.g., maintenance, slot penalties, crew legality) makes it adaptable to real-world AOCC environments.

Future work will explore deeper integration of passenger-centric objectives, real-time data assimilation, and hybrid AI optimization strategies to further enhance responsiveness and decision support capabilities. Overall, AIRS.ACR represents a significant step toward more resilient, automated, and intelligent airline disruption management.

\section*{Acknowledgements}
Quantumz.io Sp. z o.o acknowledges support received from Polish Agency for Enterprise Development (PARP), Poland under Project No. FENG.01.01-IP.02-0625/23, titled \textit{
Dynamic allocation of resources in industrial ecosystems susceptible to disturbances using physics-inspired algorithms and machine learning}.

\appendix

\section{Input/Output Data Formats}
%TODO improve explanation

\begin{itemize}[noitemsep]
    \item JSON or CSV inputs for flights, crews, slots, maintenance;
    \item Output: List of prescribed changes to the aircraft/crew/passenger schedule.
\end{itemize}

\bibliographystyle{plainnat}
\bibliography{whitepaper}
	
\end{document}